\documentclass[10pt]{wlscirep}
\usepackage[utf8]{inputenc}
\usepackage[T1]{fontenc}
\usepackage{bm}
\usepackage{lineno}
\usepackage[ruled,vlined,linesnumbered]{algorithm2e}
\usepackage[utf8]{inputenc} 
\usepackage[T1]{fontenc}    
\usepackage{hyperref}       
\usepackage{url}            
\usepackage{booktabs}       
\usepackage{amsfonts}       
\usepackage{nicefrac}       
\usepackage{microtype}      
\usepackage{lipsum}
\usepackage{fancyhdr}       
\usepackage{graphicx}       
\usepackage{amsmath}
\usepackage{color}
\usepackage{appendix}
\usepackage{xr}
\usepackage{caption}
\graphicspath{{media/}}     
\newtheorem{definition}{Definition}%
\title{A universal meta-heuristic framework for influence maximization in hypergraphs}

\author[1]{Ming Xie}
\author[1,2*]{Xiu-Xiu Zhan}
\author[1]{Chuang Liu}
\author[2]{Zi-Ke Zhang}
\affil[1]{Research Center for Complexity Sciences, Hangzhou Normal University, Hangzhou, Zhejiang, P. R. China}
\affil[2]{College of Media and International Culture, Zhejiang University, Hangzhou, Zhejiang, P. R. China}

\affil[*]{E-mail: zhanxiuxiu@hznu.edu.cn}

\begin{document}


\begin{abstract}
Influence maximization (IM) aims to select a small number of nodes that are able to maximize their influence in a network and covers a wide range of applications. Despite numerous attempts to provide effective solutions in ordinary networks, higher-order interactions between entities in various real-world systems are not usually taken into account.
In this paper, we propose a versatile meta-heuristic approach, hyper genetic algorithm (HGA), to tackle the IM problem in hypergraphs, which is based on the concept of genetic evolution. Systematic validations in synthetic and empirical hypergraphs under both simple and complex contagion models indicate that HGA achieves universal and plausible performance compared to baseline methods. We explore the cause of the excellent performance of HGA through ablation studies and correlation analysis. The findings show that the solution of HGA is distinct from that of other prior methods. Moreover, a closer look at the local topological features of the seed nodes acquired by different algorithms reveals that the selection of seed nodes cannot be based on a single topological characteristic, but should involve a combination of multiple topological features to address the IM problem.
\end{abstract}

\flushbottom
\maketitle

\thispagestyle{empty}

\section*{Introduction}
Recent research in network science has shown that interactions between more than two entities are ubiquitous in real-world systems~\cite{zhang2023higher, white1986structure, benson2016higher, iacopini2019simplicial, grilli2017higher}, thus ordinary networks with dyadic interactions are inadequate to offer a comprehensive representation of these higher-order systems. A prime example of this can be seen on online social platforms, such as WhatsApp and WeChat, where a chat group is able to include multiple participants. Hypergraphs, as one of the most representative paradigms, provide a versatile tool to describe higher-order interactions by leveraging hyperedges~\cite{ mancastroppa2023hyper, battiston2020networks, antelmi2021social, contisciani2022inference}.

As one of the fundamental problems in information diffusion, the influence maximization (IM) problem that aims to identify a fixed number of \emph{seeds} that can cause maximal influence spread in a network~\cite{kempe2003maximizing, lotf2022improved, biswas2022two, kumar2022csr, chen2022adaptive} also needs to be solved by considering high-order interactions. 

With multivariate information-sharing groups and circles, people are increasingly interconnected as they communicate with each other collectivelly, leading to a plethora of dense connections in high-order systems~\cite{alvarez2021evolutionary}.  In hypergraphs, this is crystallized by the fact that nodes are involved in multiple hyperedges, resulting in relatively high link densities and clustering coefficients in these datasets (see Methods and Materials, with most hypergraphs having a link density greater than $0.1$ and a clustering coefficient greater than $0.5$). Consequently, resolving the IM problem in hypergraphs is more difficult than in ordinary networks, as nodes may have a large influence overlap in hypergraphs when conducting a spreading process. Therefore, more attention should be paid to selecting nodes with low-influence overlap, thus maximizing their influence.

In order to tackle the difficulties posed by the IM problem in hypergraphs, researchers have developed heuristic approaches that take into account the topological features of hypergraphs~\cite{xie2022influence, zhang2023influence, ma2022hyper}. However, heuristic methods may result in locally optimal solutions~\cite{aghaee2021survey, zhang2022search, kromer2017guided}. To effectively search for a global optimal solution for the IM problem in hypergraphs, a more general framework is needed. Here, we propose a meta-heuristic algorithm, i.e., hyper genetic algorithm (HGA),  that considers both structural characteristics and diffusion dynamics to find the optimal seed set. In short, we study the problem of IM in hypergraphs with three main contributions. First, inspired by biological evolution~\cite{agarwal2018social, chatterjee2023novel}, HGA utilizes three designed operators, including selection, crossover, and mutation, to find a nearly optimal solution to the IM problem. Second, comparative experiments performed in both synthetic and empirical hypergraphs show that HGA surpasses other state-of-the-art benchmarks in terms of effectiveness and efficiency for both simple and complex contagions, suggesting a universal performance of HGA. Third, a comparison of results shows that the seeds produced by HGA are more effective due to their consideration of a wider range of characteristics, rather than just a single topological feature.

\section*{Results}\label{sec2}
{\textbf{Pipeline of Hyper Genetic Algorithm (HGA).} An efficient general framework, Hyper Genetic Algorithm, is developed to seek seed nodes for the IM problem in a hypergraph. A hypergraph provides a reliable abstraction of a complex system by encoding the multi-interactions among entities as hyperedges~\cite{ferraz2023multistability, eriksson2021choosing, young2021hypergraph}. 
Given a hypergraph $H(V, E)$, sets $V=\left\{ v_1, v_2, \cdots, v_N \right\}$ and $E=\left\{ e_1, e_2, \cdots, e_M \right\}$ denote sets of nodes and hyperedges, respectively. A hyperedge $e_j$ symbolizes the connection between $| e_j |$ nodes, where $ | e_j |$ is the number of elements in $e_j$. The connection between the nodes and the hyperedges is represented in an $N \times M$ incidence matrix, denoted as $B$. The value of $B_{i \alpha}$ is 1 if node $v_i$ is part of the hyperedge $e_{\alpha}$, and 0 otherwise. The adjacency matrix of $H$ is denoted by $A$, with each element $A_{ij}$ representing the number of hyperedges that contain both $v_i$ and $v_j$. Note that the value of the diagonal element $A_{ii}$ is set to 0.

HGA utilizes evolutionary biological patterns to identify influential nodes, as demonstrated in Fig.~\ref{fig:framework}. It is composed of a step of initialization and three evolutionary processes, namely selection, crossover, and mutation, which are briefly described below (see Methods and Materials and Supplementary Note 1 in the Supplementary Information, for a detailed description of HGA). 

The step of initialization generates an initial population that contains a set of individuals, where each individual contains $K$ non-repetitive nodes (Fig.~\ref{fig:framework}a shows the case of $K=6$). As shown in Fig.~\ref{fig:framework}b, the selection process selects individuals with high fitness scores (as measured by the fitness function given in Methods and Materials) from the initial population $\textbf{X}$, and all selected individuals then form a new population, i.e., $\textbf{X}^{'}$.
Subsequently, a crossover process is conducted. In this operation, two individuals (i.e., $X^{'}_i$ and $X^{'}_p$) are selected iteratively from $\textbf{X}^{'}$, and then nodes in $X^{'}_i$ and $X^{'}_p$ are crossed. Concretely, nodes in $X^{'}_i$ and $X^{'}_p$ are merged into $X_\text{join}$, and we sort the nodes in $X_\text{join}$ to form the sorted sequence $X^{*}_\text{join}$ (see Methods and Materials). Elite nodes in $X^{*}_\text{join}$ are retained to form the offspring. After that, mutation occurs with a given probability, which allows the evolutionary processes to find a globally optimal solution. To be specific, nodes in offspring are replaced by another one ($\hat{v}$) in the node set $V$ of the hypergraph. Note that no repetitive node in an individual should be allowed (see Fig.~\ref{fig:framework}d).
After the process of crossover and mutation, new offsprings are produced and inserted into the current selected population. In the next step of the evolutionary processes, we consider this inserted population as the input and re-conduct the selection process. As the evolutionary process iterates, the individuals in the population will eventually converge, and the converged individual is the set of seed nodes we obtain (see Methods and Materials for detailed procedures). 

\begin{figure*}[t!] 
\centering		
\includegraphics[width=\textwidth]{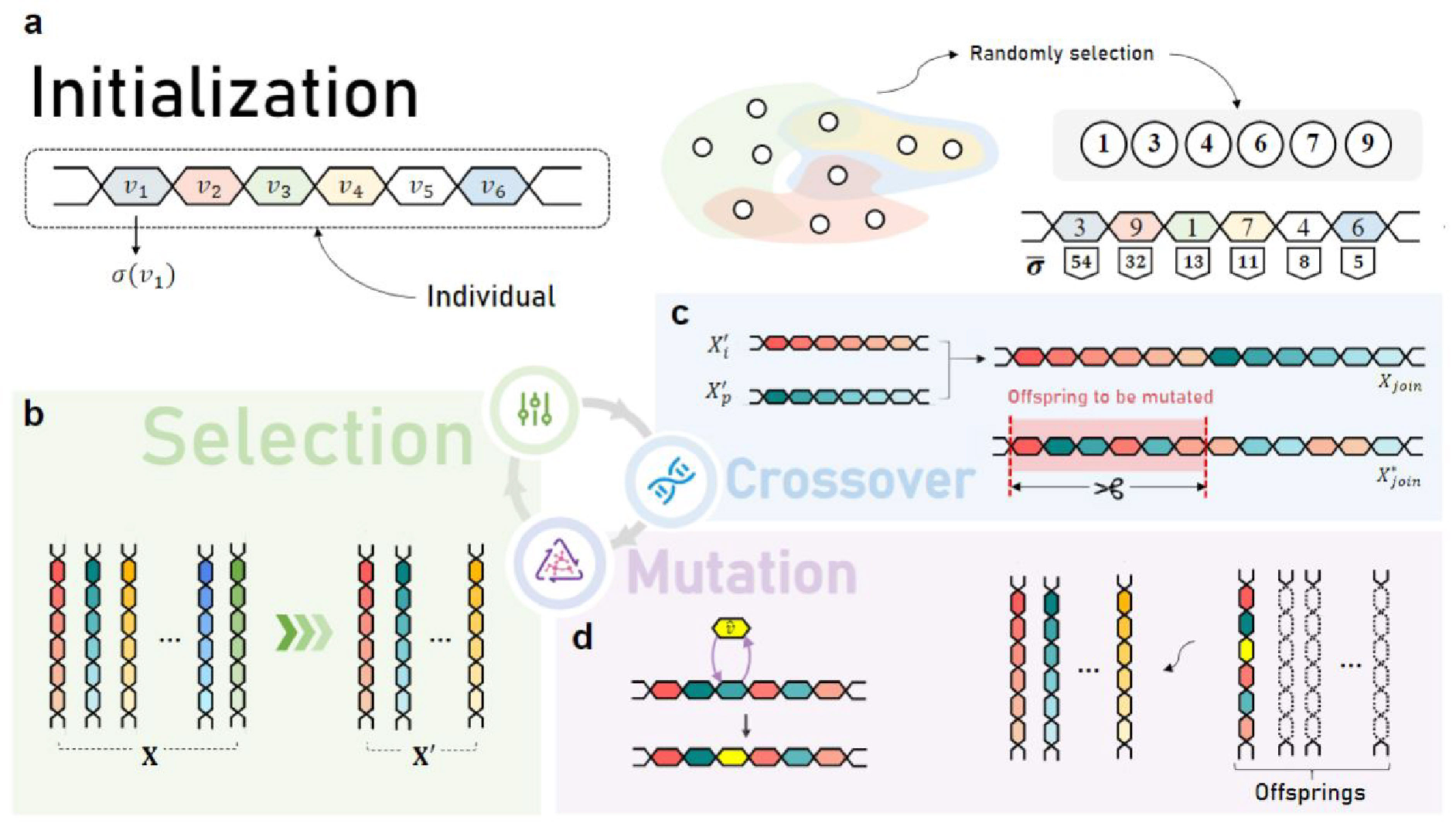}	
\caption{\textbf{Illustration of Hyper Genetic Algorithm.} \textbf{(a)} The initial step involves randomly selecting $K=6$ nodes from the set $V$ to form an individual. A population $\textbf{X}$ is then created by combining multiple individuals. \textbf{(b)}  An illustration of selection. The selection process picks individuals with high fitness scores from the pool of potential candidates in $\textbf{X}$ using the "roulette wheel" algorithm, and the chosen individuals form a new population $\textbf{X}'$. The darker the shade of the node in the diagram, the higher the value of $\overline{\sigma}$ it has. \textbf{(c)} The operator of crossover. Two parent individuals, $X'_{i}$ and $X'_{p}$, from $\textbf{X}'$ are merged by the crossover operator to form $X_\text{join}$. The nodes in $X_\text{join}$ are then sorted in descending order according to the value of $\overline{\sigma}$. The sorted ranking is referred to as $X^*_\text{join}$, and the top $K$ nodes in $X^*_\text{join}$ form a new offspring.  \textbf{(d)} The mutation process. For each offspring generated in (c), a random internal node at position $i$ is altered to a randomly chosen node  $\hat{v}$ with a fixed probability of mutation $p_m$, resulting in the evolved offspring. In our example, the random position is $i=3$. At a certain evolutionary stage, after selecting the parent individuals in a repetitive manner and going through the steps (b)-(d), the evolved offspring should be added to $\textbf{X}'$ as the potential population (i.e., $\textbf{X}$) for the following evolutionary steps.}
\label{fig:framework}
\end{figure*}

\textbf{IM in hypergraphs.}
IM in a hypergraph aims to find a seed set of nodes $S$ of size $K$ that can maximize their influence spread, i.e., the average number of nodes influenced~\cite{li2023influence, chen2016robust, banerjee2020survey} ($\sigma(S)$), for a specific spreading mechanism in a hypergraph. We propose to use two different kinds of spreading models, that is, Suspected-Infected (SI) and Linear Threshold model (LT) (see Supplementary Note 4 for detailed results of LT model), in hypergraphs, and the details of these two models are given in Sec. "Spreading dynamics" in Methods and Materials.
We assess the performance of HGA by utilizing several state-of-the-art methodologies, including HADP, HSDP, HIMR, DG, HDG, HCI and HRIS (see Methods and Materials). We compare HGA with these methodologies in terms of $\sigma(S)$. With the variation in seed size $K$, the value of $\sigma(S)$ also changes, as shown by the curve in Fig.~\ref{fig:pfm_sub}a. For a specific algorithm $\mathcal{A}$, we compute the area under the influence spread curve and name it $\Psi_{\mathcal{A}}$.  The normalized value of $\Psi_{\mathcal{A}}$ is represented as $\overline{\Psi}_{\mathcal{A}}$, and is obtained by dividing the sum of all $\Psi_{\mathcal{A}}$s computed by all algorithms' $\Psi_{\mathcal{A}}$.
A higher value of $\overline{\Psi}_{\mathcal{A}}$ implies a higher performance of the algorithm $\mathcal{A}$.

We first compare HGA with current leading baselines in synthetic hypergraphs generated by a random generator, HyperCL, which generates hypergraphs with degree distributions following different power exponents $\gamma$~\cite{xie2022influence, do2020structural}. The degree distribution's heterogeneity is inversely proportional to the exponent $\gamma$. The variation of $\sigma(S)$ with the seed set size $K$ is shown in Fig.~\ref{fig:pfm_sub}a for hypergraphs created by HyperCL with different exponents and using different algorithms. Accordingly, the comparison of $\overline{\Psi}$ is given in Fig.~\ref{fig:pfm_sub}b. HGA outperforms all other algorithms except Greedy (GD), which is approximately the optimal solution of IM with extremely high computing cost (Fig.~\ref{fig:pfm_sub}c). The time cost of the algorithms is given in Supplementary Table~\ref{tbl:TC_SYN}. Although HGA requires more time than other heuristic benchmarks, its time cost is almost 12 times less than GD (see Fig.~\ref{fig:pfm_sub}c).
In terms of $\overline{\Psi}$, it exceeds other benchmarks by more than $14.1\%$, with the exception of GD. As the value of $\gamma$ rises, HGA is still the algorithm with the largest influence spread, regardless of the seed set size (Fig.~\ref{fig:pfm_sub}a), suggesting that HGA is relatively stable when the degree of heterogeneity changes. Importantly, as $\gamma$ increases from 2.1 to 3.0, the difference between $\overline{\Psi}_{\text{HGA}}$ and the least value of $\overline{\Psi}$ (the grey area in Fig.~\ref{fig:pfm_sub}b) decreases from 0.06 to 0.015, indicating that HGA is effective in hypergraphs with high degree heterogeneity.

\begin{figure*}[t!] 
\centering		
\includegraphics[width=\textwidth]{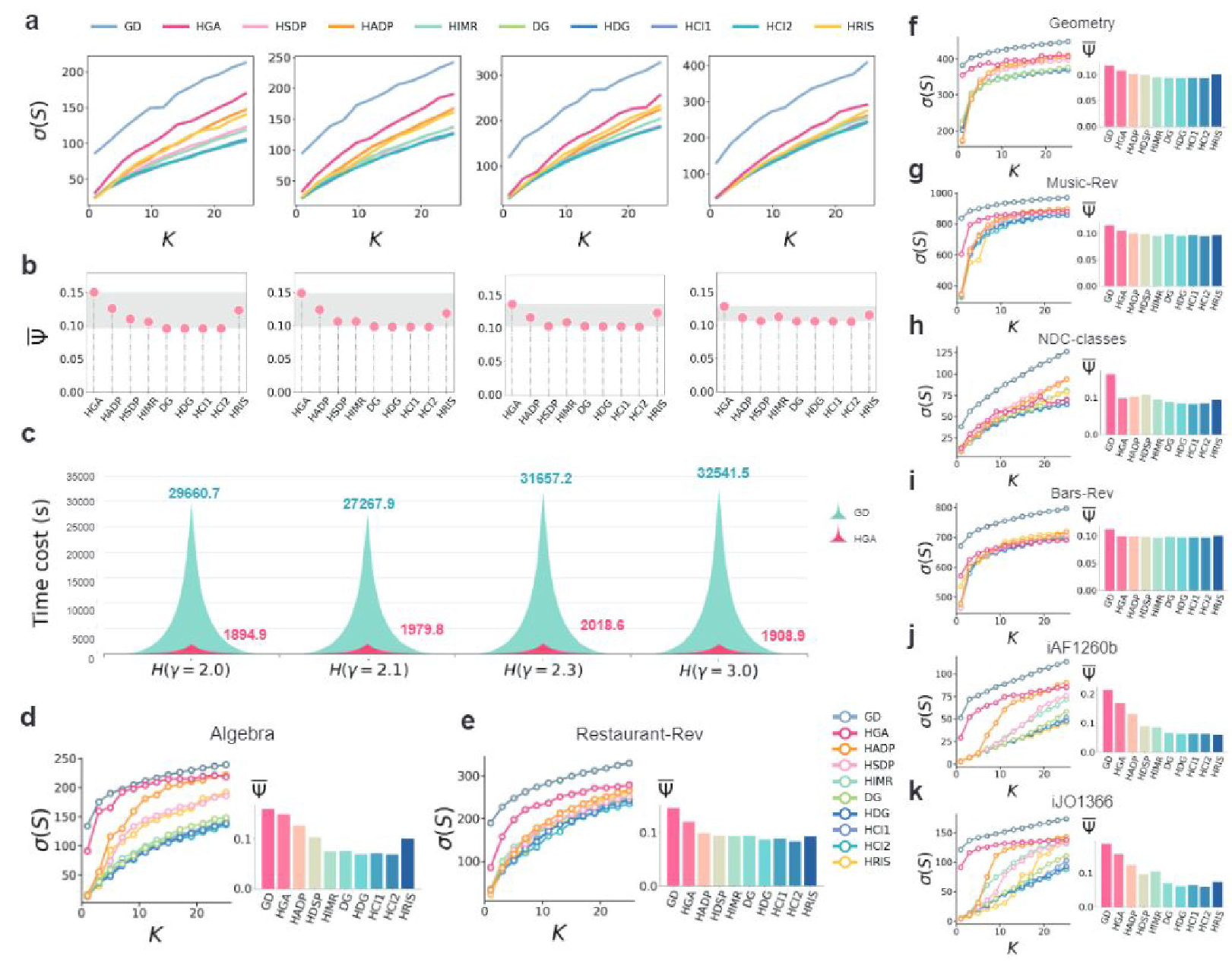}	
\caption{\textbf{The universal performance of HGA for IM problem on synthetic and empirical hypergraphs.} \textbf{(a)} The average number of influenced nodes ($\sigma(S)$) obtained by different algorithms in synthetic hypergraphs created by HyperCL, with $\gamma$ set to $2.0, 2.1, 2.3$, and $3.0$, is shown as a function of seed size $K$. \textbf{(b)} The corresponding values of $\overline{\Psi}$ for the algorithms of (a). \textbf{(c)} The comparison of time cost comparison for HGA and GD. \textbf{(d)-(k)} The average number of influenced nodes ($\sigma(S)$) as a function of seed set size $K$ in empirical hypergraphs: Algebra; Restaurant-Rev; Geometry; Music-Rev; NDC-classes; Bars-Rev; iAF1260b and iJO1366. The experiments yield results when the SI-based spreading model is used with $\beta=0.01$  and $T=25$. Each $\sigma(S)$ is obtained by the average over $500$ realizations. We treat the evolutionary step and population size in HGA as $10$ and $500$ to conduct the experiment (see Supplementary Fig.~\ref{fig:gtime} for details). }
\label{fig:pfm_sub}
\end{figure*}

Further experiments are performed in hypergraphs generated by empirical datasets from various domains, and the datasets are Algebra, Geometry, Restaurants-Rev, Music-Rev, NDC-classes, Bars-Rev, iAF1260b and iJO1366 (see Methods and Materials for a detailed description). We show the performance of each algorithm in Fig.~\ref{fig:pfm_sub}(d-k). Similar to the results of synthetic data, GD provides the best approximation of the optimal solution of IM whereas HGA performs the second best with much lower cost (see Supplementary Table~\ref{tbl:TC}). It is noteworthy that HGA exhibits similar performance to that of GD in hyper-networks such as Algebra, Restaurants-Rev, and iAF1260b. Specifically, it surpasses the HADP with a maximal $28.6\%$ improvement in iAF1260b. The algorithms HADP, HSDP, and DG are all based on degrees, and HADP has superior performance compared to the other two. This is because its internal adaptive penalty mechanism eliminates more redundant influence overlap as the value of $K$ increases. For the hyperdegree-based methods, i.e., HCI and HDG, their performance is much worse than the above ones. We observe that HCI outperforms HDG, with an improvement of at least $0.21\%$. This indicates that taking into account the hyperdegree of distant nodes is beneficial for identifying influential nodes. HRIS is not as effective as HGA, yet it surpasses those techniques based on hyperdegree in the majority of hypergraphs. This could suggest that the impact of a node can be estimated by its frequency in reverse reachable sets. However, the randomness of the hyperedge pruning process may cause its performance to be inconsistent. Overall, it appears that HGA is highly effective when applied to both synthetic and real-world hypergraphs, indicating that evolutionary techniques for selecting seed nodes are more successful than heuristic-based approaches in hypergraphs.

\textbf{Ablation analysis.} We further investigate the driving force of HGA that leads to its high performance by analyzing the parameters of the crossover and mutation operators.
As an illustration, Fig.~\ref{fig:subplots}a shows how $\overline{\Psi}$ fluctuates when both the probability of crossover $p_c$ and the evolutionary step are altered in Algebra. The results for other hypergraphs are given in Sec. "Ablation analysis" in the Supplementary Information. The increase of the crossover probability $p_c$ $(p_c \in [0, 1])$ implies an enhanced role of the crossover mechanism. Given a specific evolutionary step, the effectiveness of HGA improves as $p_c$ increases in most hypergraphs. To be specific, the improvement of $\overline{\Psi}$ of $p_c=1$ is $26.90\%$ compared to that of $p_c=0$ for Algebra when we fix the evolutionary step as $9$. The reason is that high crossover probability enables the parental individuals to produce new offsprings with high fitness scores. In addition, 
as the evolutionary step grows, the effectiveness of the algorithm also improves significantly for any fixed value of $p_c$. This is attributed to the fact that a larger evolutionary step leads to more elite nodes being selected, and thus the evolutionary process can converge to the optimum more rapidly. Supplementary Fig.~\ref{fig:ablation} also shows consistent results.

With respect to the mutation process, the probability of mutation $p_m$ is usually set to a relatively small number. Therefore, we explore the values of $\sigma(S)$ with $p_m$ equal to $0$ (that is, without considering the mutation process), $0.005$, and $0.01$. As shown in Fig.~\ref{fig:subplots}b, we observe that the performance of HGA with $p_m=0.005$ is slightly better than that of $p_m=0$. Additionally, we find that $\sigma(S)$ is not stable for different seed set sizes when $p_m=0.01$. Therefore, we use $p_m=0.005$ in our study.

In brief, the crossover operator is seen as the main factor behind the success of the HGA. The mutation process also contributes, but it may not have a major impact when the evolutionary step is already saturated.

\begin{figure*}[t!] 
\centering		
\includegraphics[width=\textwidth]{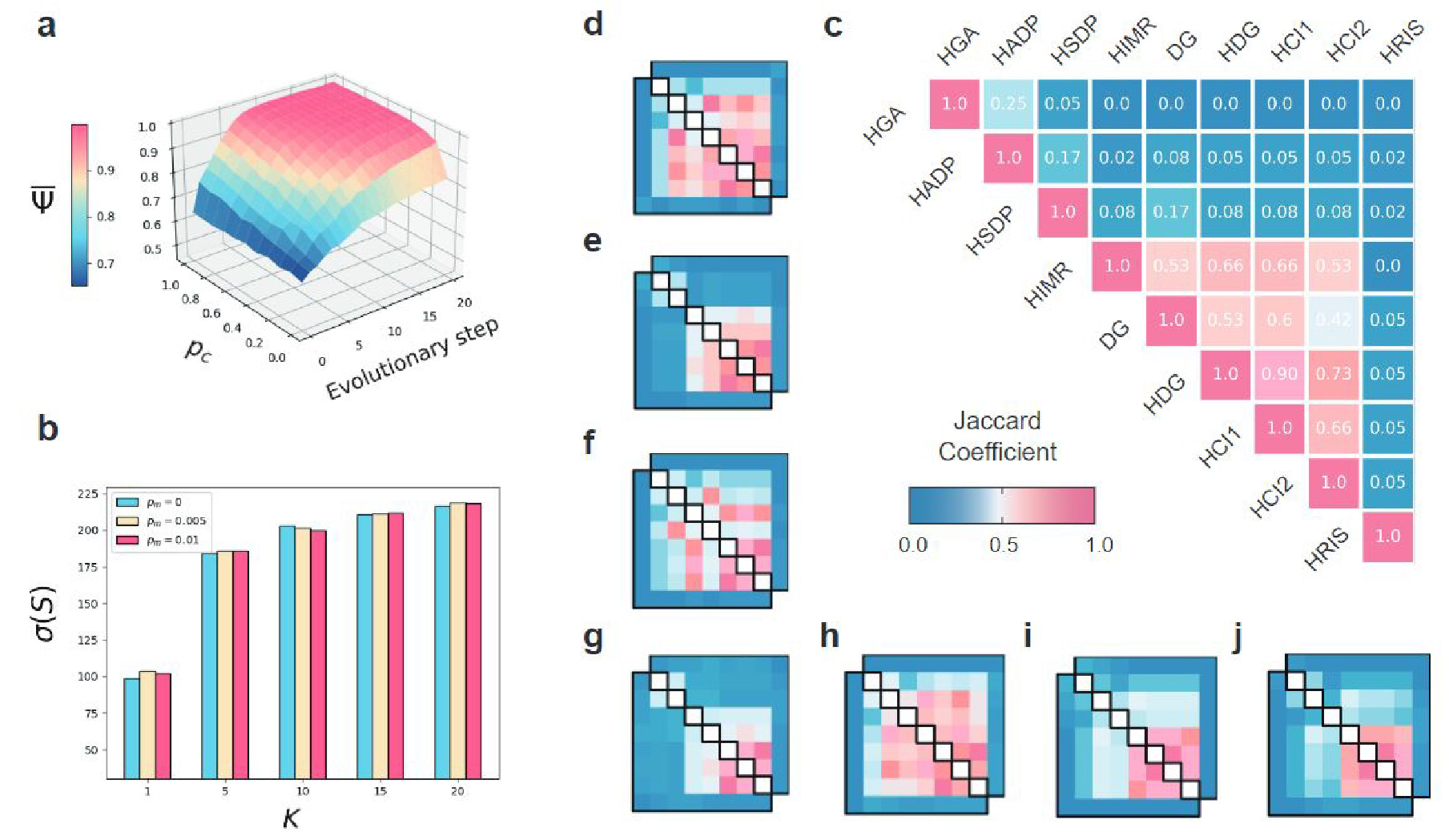}	
\caption{\textbf{Ablation studies and correlation analyses.} \textbf{(a)} 
The values of $\overline{\Psi}$ for HGA vary depending on the evolutionary step and the crossover probability $p_c$. 
Each value of $\overline{\Psi}$ for a specific pair of evolutionary step and $p_c$ is calculated by the average over 10 times. \textbf{(b)} The average value of $\sigma(S)$ is shown as a function of the seed set size $K$ for different mutation probabilities $p_m$. For each fixed $K$, the influence spread of HGA with $p_m$ of $0, 0.005$, and $0.01$, respectively, is represented by bars of different colors, i.e., blue, yellow, and pink. \textbf{(c)-(j)} Correlation between seed sets obtained by different algorithms, we show the results for the following hypergraphs: Algebra, Restaurant-Rev, Geometry, Music-Rev, NDC-classes, Bars-Rev, iAF1260b and iJO1366. 
When two sets of seeds generated by different algorithms have a high Jaccard coefficient, the value of the related element in the matrix is high (shown in pink), and vice versa (shown in blue). We present the outcomes for a seed set size of 20. Correlation analysis in Supplementary Information contains the results for other seed set sizes, which are in agreement with the findings here.}
\label{fig:subplots}  
\end{figure*}

\textbf{Comprehensive topological features are required.} To comprehend why HGA is more successful than the other algorithms, we investigate the characteristics of the seed sets acquired by the various algorithms. First, we utilize the Jaccard index (see Methods and Materials for details) to measure the similarity between the seed sets obtained by any two algorithms. The similarity matrices of various empirical hypergraphs are displayed in Fig.~\ref{fig:subplots}(c-j). Taking Fig.~\ref{fig:subplots}c as an example (the hypergraph of Algebra), the seed set obtained by HGA is dramatically different from any of the baselines, i.e., most of the Jaccard similarity is $0$. This
means that our algorithm jumps out of the shackles of previous algorithms and thus shows a greater spread of influence. Alternatively, the seed sets obtained by degree- and hyperdegree-based algorithms such as HCI, HDG, and DG are quite similar, suggesting that degree-based methods are highly correlated with hyperdegree-based ones, which is consistent with previous findings~\cite{xie2022influence}. Meanwhile, adaptive degree-based methods, such as HADP and HSDP, tend to have a lower similarity to other algorithms because of their degree pruning process when selecting seed nodes. Moreover, HRIS shows a small similarity with other benchmarks, which can be traced back to its internal randomness within the hyperedge pruning process.

\begin{figure*}[t!] 
\centering		
\includegraphics[width=\textwidth]{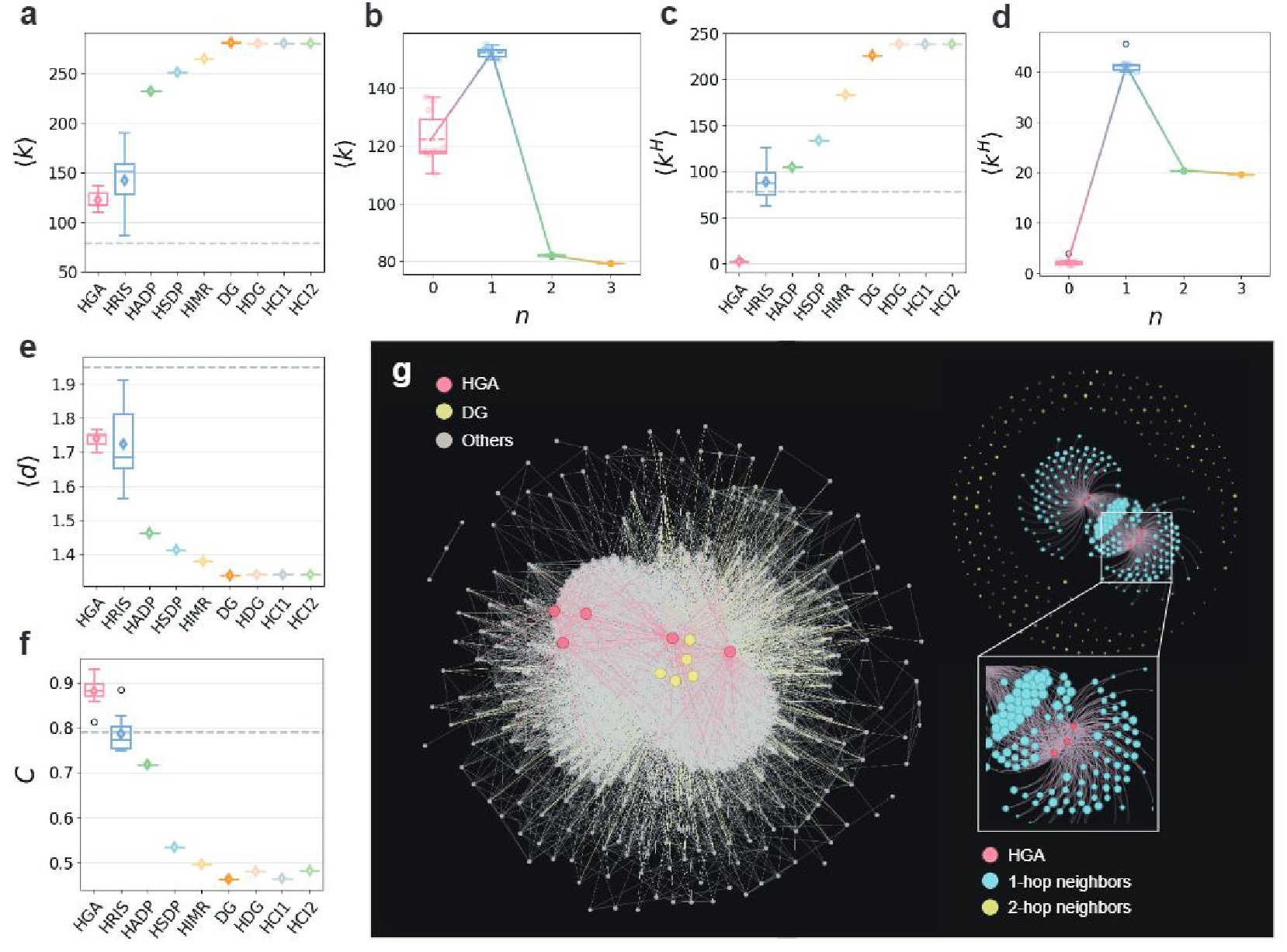}	
\caption{\textbf{Topological analyses and visualizations of seed nodes in Algebra.} \textbf{(a)} Comparison of the average degree ($\left\langle k \right\rangle$) of the seed nodes obtained by different algorithms. The median and mean values in the boxes of (a), (c), (e), and (f) are represented by a line and a rhombic, respectively. \textbf{(b)} Average degree ($\left\langle k \right\rangle$) of the $n$th-order ($n \in \{0,1,2,3\}$) neighbors of the seed nodes selected by HGA. The value $n=0$ indicates the seeds directly. The dashed and the solid lines represent the mean and the median, respectively, for the boxes in (b) and (d). \textbf{(c)} Average hyperdegree ($\left\langle k^H \right\rangle$) of the seed nodes selected by different algorithms. \textbf{(d)} Average hyperdegree $\left\langle k^H \right\rangle$ of the seeds' $n$th-order neighbors, the seeds are obtained by HGA. \textbf{(e)} Average distance ($\left\langle d \right\rangle$) between the seeds obtained by different algorithms. \textbf{(f)} Average clustering coefficient ($C$) of the seed nodes identified by different algorithms. \textbf{(g)} Visualization of the seeds obtained by  HGA (pink) and DG (yellow) (left panel). We note that the clique expansion of the original hypergraph is employed. Also, we show the local structure of HGA seed nodes, as shown in the upper right panel. The pink dots denote the HGA seed nodes, and their first- and second-order neighbors are plotted in the color blue and yellow, respectively. Pink links represent the interactions between the seeds and their first-order neighbors. For clearness, we omit the connection between the first- and second-order neighbors in the figure. The size of nodes is related to the node degree of the original hypergraph, with larger nodes indicating a higher degree. Each value of the measurement for (a)-(f) is obtained over 10 realizations in Algebra with seed size $K=5$ (results for other $K$ values and other hypergraphs are depicted in Supplementary Fig.~\ref{fig:visualization_si_1}-~\ref{fig:visualization_si_2}.}
\label{fig:visualization}
\end{figure*}

In Fig.~\ref{fig:visualization}, we show the topological characteristics of the seeds and their $n$-order neighbors of hypergraph Algebra, where the seed size $K=5$. The other hypergraphs and seed sizes show patterns similar to those in Fig.~\ref{fig:visualization} and are given in Supplementary Fig.~\ref{fig:topo_deg_K_5}-~\ref{fig:topo_cls_K_20}. Fig.~\ref{fig:visualization}a and c illustrate the results of average node degree $\langle k \rangle$ and node hyperdegree $\langle k^{H} \rangle$, respectively. The average node degree (or hyperdegree) of seed sets obtained by HGA is much lower than that of the other algorithms, especially much lower than that of the degree-based IM algorithms. In other words, it indicates that high-degree (or high-hyperdegree) nodes may not necessarily be selected as seeds for the IM problem. Fig.~\ref{fig:visualization}b (or d) shows the average degree (or hyperdegree) of the $n$-order neighbors of the seed nodes obtained by HGA. It is worth mentioning that $n=0$ indicates the average degree (or hyperdegree) of the seeds. With increasing $n$, the highest average degree (or hyperdegree) appears when $n=1$, suggesting that seeds tend to be directly connected to nodes with high degree (or hyperdegree) while the $n$-order neighbors ($n>1$) tend to be marginal nodes with relatively low degree (or hyperdegree). We compared the average distance between seeds obtained by different algorithms in Fig.~\ref{fig:visualization}e, where the distance between nodes is calculated based on the clique expansion of the hypergraph. One can observe that the average distance between the seeds of HGA is larger than that of the other algorithms. As a matter of fact, distant nodes tend to have a low probability of influence overlap when selected as seeds. This means that the large distance between the seeds of HGA could be a possible reason for its high performance in solving the IM problem. Similarly, we also explore the average hypergraph clustering coefficient~\cite{zeng2023hyper} of seed nodes, i.e., the ratio of the number of existing hyperedges between its neighbors to the number of the possible hyperedges, the results are given in Fig.~\ref{fig:visualization}f. HGA seeds display a higher average clustering coefficient than the others, meaning that the seeds may have a denser local structure than those of the other algorithms. 

To clarify the insights observed above, we visualize the seeds of HGA and compare them with those of DG in Fig.~\ref{fig:visualization}g. There is no overlap between HGA seeds and DG seeds, and the HGA seeds are more distant than those of DG. The local structure of the HGA seeds is further displayed in the right panel of Fig.~\ref{fig:visualization}g. The blue and yellow nodes are the first- and second-order neighbors of HGA seeds. The degrees of them are represented by the size of the nodes. The figure shows that HGA seeds are directly connected to high-degree nodes, while their second-order neighbors have relatively lower degrees. We claim that the results are consistent with the analysis given in Fig.~\ref{fig:visualization}a-f. And such a large difference in the topological features of the seeds may together make HGA perform better than the others for IM.

\section*{Discussion}\label{sec12}
The main challenge of IM is to develop a relatively low time cost, yet well-performing algorithm to find $K$ seeds that can maximize spread influence in a network, which has been extensively studied. In this paper, we develop a universal meta-heuristic solution, Hyper Genetic Algorithm (HGA), to solve the IM problem in a hypergraph. In HGA, the initialization phase and the three genetic operators are combined to create an effective framework for selecting seed nodes. Initially, a population with a fixed amount of individuals is constructed. The original population is filtered through a selection operator which chooses individuals with the highest fitness scores, as determined by a pre-defined fitness function. To ensure the diversity of individuals in the population, we design a reorganization mechanism through the crossover operation. Afterwards, a mutation operator is used to prevent the algorithm from becoming stuck in a local optimal solution. After iteratively updating the individuals in the population, the individuals eventually converge, and we obtain the solution for the selection of seed nodes.

The effectiveness of HGA is validated in hypergraphs from various domains, including synthetic and empirical ones. The result of synthetic hypergraphs implies strong applicability of our solution, especially in hypergraphs with high degree heterogeneity, which is ubiquitous in real-world systems. The adaptive fitness function designed in HGA makes it possible to be applied to different spreading models, and we verified the applicability using two classical spreading models, SI and LT model, which mimics the simple and complex contagion, respectively.
Further exploration of the driving force of HGA reveals that two operators, i.e., crossover and mutation,  especially the crossover operator, provide basic guarantees for its remarkable performance.

We then investigate the seeds obtained by different algorithms by looking at the overlap between different seeds. Experimental results show that most of the previous approaches show high overlaps, whereas our algorithm finds seeds that are quite different from other methods. It may suggest that our algorithm jumps out of the previous trap of selecting seed nodes. The observations obtained above further inspire us to deeply research the properties of the seeds obtained by HGA. Quantitative analysis and visualization of seeds show that HGA seeds possess a lower average degree, a lower average hyperdegree, a larger distance between each other, and a higher average clustering coefficient than those of the other baselines. The findings may suggest that the solution for the IM problem in a hypergraph cannot be based only on a specific topology property, such as degree, hyperdegree, etc. Instead, a comprehensive metric that incorporates various topological properties should be proposed, which requires further exploration in the future. 
Last but not least, there are many open problems for IM in a hypergraph that deserve future attention. For example, the question of how to design a generic fitness function using influence estimation for the sake of time cost of evolutionary algorithm~\cite{gong2016influence}. In addition, solutions to targeting the influential seeds in large-scale hypergraphs are also worth pursuing. As different types of higher-order networks are revealed in various domains, we expect to find extended solutions in other higher-order network representations, such as simplicial complexes~\cite{wang2022full, fan2021characterizing}.

\section*{Methods and materials}\label{sec11}

\subsection*{Procedures in Hyper Genetic Algorithm}\label{HGA}
\textit{Initialization.} We begin HGA by generating an initial population $\textbf{X}=\{X_1, X_2, \cdots, X_{\eta*l_p}\}$. Each individual $X_i=(v_{i_1}, v_{i_2}, \cdots, v_{i_K})$ is composed of $K$ distinct nodes randomly chosen from $V$. The parameter $\eta$ (which is greater than 1) is used to adjust the size of the initial population, and $l_p$ is the size of the population after the selection process in the following evolutionary steps. 

\textit{Selection.} Selection aims to select individuals with high fitness scores from the pool of potential candidates $\textbf{X}$, which is analogous to the evolutionary law that states those adaptable creatures are more likely to survive. 
Specifically, at the selection stage, each element $X_i$ in the set $\textbf{X}$ is assigned a fitness score $f(X_i)$ by means of a fitness function. We choose $l_p$ individuals from $\textbf{X}$ based on the roulette wheel algorithm~\cite{mirjalili2019genetic}, and then the individuals with relatively high fitness scores are selected to form a new population, that is, $\textbf{X}'$.

To introduce the selection process, we begin with the definition of the fitness function $f(\cdot)$, which is used to generate the fitness score for each individual.
\begin{definition} [Average influence spread of a node]
\label{Average influence spread of node} Given a specific spreading model, the total number of infected nodes at time $T$ starting from node $v$ is denoted $\sigma(v)$, which is called influence spread of $v$. 
If we repeat the spreading from $v$ for $R$ times, we can get the influence spread from node $v$ as $\sigma^{r}(v), 1\leq r \leq R$. 
Therefore, the average influence spread of $v$ is given by the following equation:

\begin{equation}
\overline{\sigma}(v) = \frac{1}{R}\sum_{r=1}^{R}\sigma^{r}(v),
\end{equation}
\end{definition}

\begin{definition} [Fitness function]
\label{Fitness function} Given an individual $X_i=(v_{i_1}, v_{i_2}, \cdots, v_{i_K})$, the fitness score $f(X_i)$ of $X_i$ is defined by the sum of the average influence spread of every node in $X_i$.  Mathematically speaking, $f(X_i)$ is represented as
\begin{equation}
f(X_i) =  \sum_{j=1}^{K}\overline{\sigma}(v_{i_j}).
\end{equation}

\end{definition}

We then use the "roulette wheel" mechanism to select individuals with high fitness scores from a population $\textbf{X}$. Concretely, the ratio of the fitness score of $X_i$ to the total score of all individuals in the candidate population, i.e., ${f(X_i)}/{\sum_jf(X_j)}$, is calculated for $X_i$. The ratio represents the probability of $X_i$ being selected during the "roulette wheel" mechanism, i.e., the larger the value, the more likely $X_i$ can be selected.
Taking the candidate population $\textbf{X}$ in Fig.~\ref{fig:framework} as an example, $l_p$ individuals in $\textbf{X}$ are selected into a new population $\textbf{X}'$ through the "roulette wheel" mechanism. The pseudocode of the selection stage is shown in the Supplementary Algorithm~\ref{selection}.

\textit{Crossover and Mutation.}
Subsequently, the crossover and mutation operators of the algorithm begin to take effect.
In the crossover, we iteratively take a pair of individuals from $\textbf{X}'$ with probability $p_c$ and consider them as parents. To be specific, we generate a random number $z\in[0, 1]$ for each individual $X'_{i}$ in $\textbf{X}'$. If $z<p_c$, we randomly select one individual $X_{p}'$ from $\textbf{X}'$.
Then these parental individuals are merged into a long one, that is, $X_{\text{join}}$. As shown in Fig.~\ref{fig:framework}, nodes in $X_{\text{join}}$ are sorted in descending order according to their average influence spread $\overline{\sigma}$, and the new merged individual is named $X^{*}_{\text{join}}$. The top $K$ nodes with the highest $\overline{\sigma}$ are further selected to form an offspring of $X^{*}_{\text{join}}$ and will be used in the mutation phase with a probability $p_m$. However, if $z\geq p_c$, we assign $\hat{X}=X'_{i}$. In the mutation, a random node of the offspring has a probability to be replaced by another one from $V$, and then an evolved offspring is finally obtained. Specifically, mutation allows a random position in an offspring to mutate to another node with a probability $p_m$. In particular, for the offspring $\hat{X}$, an internal node in a random position $i$ should be replaced by another random chosen node $\hat{v}$ ($\hat{v} \in V \backslash \hat{X}$) with probability $p_m$.
Likewise, we ensure that there are no duplicate nodes in an individual during the mutation stage. Then the final offspring $\tilde{X}$ is obtained and added to the population $\textbf{X}'$.  
The newly evolved offsprings obtained by the current iteration of evolutionary procedures are then squeezed into the
population, and then HGA comes to the selection stage in the next iteration again. Iteration of the evolutionary procedures is stopped when the evolutionary step reaches a certain point that causes the node sequence $X$ to converge. The individual that has converged is the set of seed nodes that has been selected by HGA.

The pseudocodes of the operators of crossover and mutation are shown in the Supplementary Algorithm~\ref{crossover} and the Algorithm~\ref{mutation}, respectively.

\subsection*{Spreading dynamics}\label{SD}
We introduce two spreading models, namely the SI model with the collective process and the Linear Threshold model to mimic the simple and complex contagion process in a hypergraph~\cite{horsevad2022transition}, respectively. 

\textit{Susceptible-Infected (SI) spreading model with collective process~\cite{xie2022influence}.}  In this model, an individual can only take one of the two states, i.e., susceptible (S) or infected (I). The set of seed nodes $S_0$ is selected to be infected at time $t=0$. 
At each time step $t$, for each of the infected nodes $v_i$, we randomly choose one hyperedge $e$ that contains $v_i$. For each of the S-state nodes in $e$, it will be infected by $v_i$ with an infection probability $\beta$. The dynamic process ends when a specific time step $T$ reaches, where $T$ is a control parameter.

\textit{Linear Threshold (LT) model.} The linear threshold model is used to conduct the complex contagion process in a hypergraph\cite{xu2022dynamics, goyal2011simpath, lu2014efficient, he2019tifim}. The rationale of the LT model is that the activation state of a node is influenced by the joint action of its neighboring nodes. In this model, the active weight $\omega_{ij}$ of a node $v_i$ to activate its neighbor $v_j$ is defined as:
\begin{equation}
    \omega_{ij} = \frac{A_{ij}}{\sum_j{A_{ij}}}.
\end{equation}

Initially, we assign an activation threshold $\theta_i (\theta_i \in [0,1])$ for each node $v_i$, where $\theta_i$ is generated by a uniform distribution in the range of $[0,1]$, and the seeds are activated simultaneously. 
At time step $t(t\geq1)$, an inactive node $v_b$ will be activated if $\sum_{v_a\in \mathcal{I}_{t-1}, v_a\in\mathcal{N}_b} \omega_{ab} > \theta_b$, where $\mathcal{I}_{t-1}$ denotes the activated node set before $t$ and $\mathcal{N}_b$ represents the neighboring node set of $v_b$, that is, each node $v_a$ in $\mathcal{N}_b$ shares at least one hyperedge with $v_b$. The dynamic process is terminated when the time step reaches $T$.

\subsection*{Benchmarks}\label{BS}

We utilize state-of-the-art benchmarks to validate the effectiveness of the proposed algorithm, which are described as follows. 

We start with the simplest methods for influence maximization, i.e., the degree-based methods. First of all, the degree index (DG) chooses  top-$K$ nodes with the largest degree as the seed set $S$. The previous study shows that there is a large influence overlap of the seed nodes selected by DG~\cite{xie2022influence}. Thus, two adaptive degree-based algorithms, i.e., Hyper Adaptive Degree Pruning (HADP) and Hyper Single Degree Pruning (HSDP)~\cite{xie2022influence}, are then proposed to effectively reduce the influence overlap of the seeds. HADP and HSDP use a degree pruning mechanism that aims to give more penalties to a node that has more neighbors in $S$. It means that the neighboring nodes of $S$ have a reduced probability of becoming seeds. The difference between HADP and HSDP is that the penalty in HADP is positively correlated to the adaptive degree of a node, whereas the penalty in HSDP is a constant value.

The hyperdegree of a node (HDG) indicates the number of hyperedges to which the node belongs~\cite{ouvrard2020hypergraphs}. Similarly to DG, HDG selects the top $K$ nodes with the highest hyperdegree as seeds. Additionally, we use Hyper Collective Influence (HCI) which is based on the node hyperdegree as a benchmark.
With the definition of a $Ball$ in a hypergraph~\cite{morone2016collective, morone2015influence}, the $hci$ value of each node $v_i$ is given by $hci_l(v_i) = (k^H(v_i)-1)\sum_{v_j \in \partial Ball(v_i, l)}(k^H(v_j)-1)$, where $k^H(v_i)$ is the hyperdegree of node $v_i$ and $l$ is a tunable parameter to adjust the orders of the considered neighbors. When sorting the $hci$ value of each node in descending order, the top $K$ nodes with the highest value are selected as $S$. Note that HCI1 and HCI2 denote the algorithm HCI performed when $l=1$ and $l=2$, respectively.

Furthermore, a global method, namely Hyper Reverse Influence Sampling (HRIS), is introduced, which adopts a reverse influence sampling process in a hypergraph~\cite{sun2021influence, xie2022influence}. Specifically, hyperedges in the original hypergraph are removed by an adaptive probability and then formed a sub-hypergraph. In the sub-hypergraph, we randomly select a node $v_i$ and define a Hyper Reverse Reachable set (HRR set) of $v_i$. The HRR set of $v_i$ is constructed by a connected component started from $v_i$. We obtain $\tau$ HRR sets by repeating the procedure above $\tau$ times.
 The seeds are then iteratively selected based on the $\tau$ HRR sets. 
In each step $t$ ($t\in[1, K]$) of the selection of seed nodes, a node $v_t$ with the highest frequency in the $\tau$ HRR sets is selected into $S$, and we subsequently remove the HRR sets that contain $v_t$. The node selection iteration ends when $K$ nodes are selected.

Besides, a ranking-based method, i.e., Hyper-IMRANK (HIMR), is applied. HIMR begins with an initial ranking of node influence by node degree and aims to obtain a convergent and self-consistent node sequence by iteratively updating the node ranking based on the estimate of node marginal influence. Each iteration includes two processes, namely influence estimation and node re-ranking. The algorithm first estimates node influence via the hyper last to first allocating strategy and then sorts the nodes in descending order based on the estimated values. The iteration is stopped until the ranking converges. The top-$K$ nodes in the convergent ranking are collected as $S$.

For comparison, the greedy algorithm~\cite{kempe2003maximizing, xie2022influence} that takes into account both topological and dynamical information is performed to give a decent approximation of near-optimal influence spread and identify a quasi-optimal set of seed nodes. We denote the marginal gain of influence for node $v$ in time step $t$ as $\mathcal{G}^t_{v} = \sigma(S_{t-1}\cup \left\{ {v} \right\})-\sigma(S_{t-1})$, where $S_{t-1}$ indicates the selected seed node set before time $t$. In the algorithm, the seeds are iteratively selected based on the marginal gain of influence. At each time step $t$, node $v^*$ with the largest marginal gain of influence, that is, $v^* = \arg\max\left\{ \mathcal{G}^t_v \right\}, v\in V \backslash S_{t-1}$, is chosen as a new seed and placed in $S_t$, i.e., $S_{t} = S_{t-1} \cup \left\{ v^* \right\}$. The greedy algorithm stops when $K$ nodes are selected as seeds.

\subsection*{Jaccard similarity coefficient}\label{JC}
Jaccard similarity~\cite{lu2017link, dharavath2016entity} is a coefficient that measures the resemblance between two finite sets of samples. Given two sets containing several elements, i.e., $\xi = \left\{\xi_1, \xi_2, \cdots, \xi_K \right\}$ and $\zeta = \left\{ \zeta_1, \zeta_2, \cdots, \zeta_K \right\}$, the Jaccard similarity coefficient is the ratio of the intersection size to the union size of the two sets, which can be mathematically defined as

\begin{equation}
J(\xi, \zeta) = \frac{\lvert \xi \cap \zeta \rvert}{\lvert \xi \cup \zeta \rvert}.
\end{equation}

We note that $J(\xi, \zeta) \in [0, 1]$, and $J(\xi, \zeta)$ is close to $1$ if the two sets $\xi$ and $\zeta$ are very similar and vice versa.

\subsection*{Datasets}\label{Datasets}

\textbf{Synthetic hypergraphs.}
We apply HyperCL random generator\cite{do2020structural, xie2022influence} to generate a hypergraph based on a given hyperdegree distribution. In HyperCL, the hyperdegree sequence is generated by a hyperdegree distribution $p(k^H) \sim (k^H)^{-\gamma}$, where the exponent $\gamma$ is a tunable parameter. 
Given an empty hyperedge $e_a (e_a \in E)$ with size $|e_a|$, we add node $v_i$ to $e_a$ with probability $k^H(v_i) / \sum_{j=1}^{N}k^H(v_j)$. Note that the node addition process for each hyperedge is terminated until the size of the hyperedge reached the given value. The hypergraph generation stops when all hyperedges are fulfilled.

In HyperCL, the distributions of hyperdegree and degree change from heterogeneity to homogeneity as $\gamma$ increases, as the degree and hyperdegree are positively correlated~\cite{xie2022influence}. We choose $\gamma \in \{2.0, 2.1, 2.3, 3.0\}$ to generate synthetic hypergraphs.

We investigate the topological properties of the synthetic hypergraphs above, e.g., the number of nodes and hyperedges, the average degree of nodes, the average hyperdegree of nodes, the average size of hyperedges, the average length of the shortest path, the diameter and the link density, which are shown in Table~\ref{tab:topo_syn}. 

\begin{table*}[htb]
    \scriptsize
    \centering
        \caption{\newline Topological properties of synthetic hypergraphs. $N$ and $M$ represent the number of nodes and hyperedges in a hypergraph, respectively, $\left\langle k \right\rangle$ is the average of node degree, $\left\langle k^H \right\rangle$ is the average of node hyperdegree,  $\left\langle k^E \right\rangle$ represents the average size of the hyperedges, which is given by the number of nodes in a hyperedge. $C$, $\left\langle d \right\rangle$, $\xi$, and $\rho$ are the clustering coefficient, the average of shortest path length, diameter, and link density of the clique expansion of a hypergraph. }
        \resizebox{\textwidth}{!}{
    \begin{tabular}{cccccccccc}
    \toprule
		Hypergraphs &$N$&$M$& $\left\langle k \right\rangle$ & $\left\langle k^{\tiny H} \right\rangle$&$\left\langle k^{\tiny E} \right\rangle$&$C$&$\left\langle d \right\rangle$&$\xi$&$\rho$\\
	\midrule
		$H(\gamma=2.0)$ & $1.0 \times 10^{3}$ & $1.0 \times 10^{3}$ & 65.55 & 11.89 & 11.09 & 0.70 & 1.94 & 3 & 0.07 \\
		$H(\gamma=2.1)$ & $1.0 \times 10^{3}$ & $1.0 \times 10^{3}$ & 71.39 & 11.33 & 10.95 & 0.63 & 1.92 & 3 & 0.07 \\
		$H(\gamma=2.3)$ & $1.0 \times 10^{3}$ & $1.0 \times 10^{3}$ & 97.49 & 11.17 & 11.09 & 0.49 & 1.91 & 3 & 0.10 \\
		$H(\gamma=3.0)$ & $1.0 \times 10^{3}$ & $1.0 \times 10^{3}$ & 115.208 & 10.84 & 10.84 & 0.36 & 1.89 & 4 & 0.12 \\
	\bottomrule
    \end{tabular}
        }
    \label{tab:topo_syn}
\end{table*}

 \begin{table*}[htb]
    \scriptsize
    \centering
        \caption{\newline Topological properties of empirical datasets. $N$ and $M$ represent the number of nodes and hyperedges in a hypergraph, respectively, $\left\langle k \right\rangle$ is the average of node degree, $\left\langle k^H \right\rangle$ is the average of node hyperdegree,  $\left\langle k^E \right\rangle$ represents the average size of the hyperedges, which is given by the number of nodes in a hyperedge. $C$, $\left\langle d \right\rangle$, $\xi$, and $\rho$ are the clustering coefficient, the average of shortest path length, diameter, and link density of the corresponding clique expansion of a hypergraph. }
        \resizebox{\textwidth}{!}{
    \begin{tabular}{cccccccccc}
    \toprule
		Hypergraphs &$N$&$M$& $\left\langle k \right\rangle$ & $\left\langle k^{\tiny H} \right\rangle$&$\left\langle k^{\tiny E} \right\rangle$&$C$&$\left\langle d \right\rangle$&$\xi$&$\rho$\\
	\midrule
		Algebra & 423 & 1268 & 78.90 & 19.53 & 6.52 & 0.79 & 1.95 & 5 & 0.19 \\
		Restaurant-Rev & 565 & 601& 79.75 & 8.14 & 7.66 & 0.54 & 1.98 & 5 & 0.14 \\
		Geometry & 580 & 1193& 164.79 & 21.53 & 10.47 & 0.82 & 1.75 & 4 & 0.28 \\
		Music-Rev & 1106 & 694 & 167.87 & 9.49 & 15.13 & 0.62 & 1.99 & 8 & 0.15 \\
		NDC-classes & 1161 & 1088 & 10.71 & 5.55 & 5.92 & 0.61 & 3.50 & 9 & 0.01 \\
		Bars-Rev & 1234 & 1194 & 174.30 & 9.61 & 9.93 & 0.58 & 2.10 & 6 & 0.14 \\
		iAF1260b & 1668 & 2351 & 13.26 & 5.46 & 3.87 & 0.55 & 2.67 & 7 & 0.007 \\
		iJO1366 & 1805 & 2546 & 16.91 & 5.55 & 3.94 & 0.58 & 2.62 & 7 & 0.009 \\
	\bottomrule
    \end{tabular}
        }
    \label{tab:topo}
\end{table*}

\textbf{Real-world datasets.} \textit{Algebra}~\cite{amburg2020fair} and \textit{Geometry}~\cite{amburg2020fair} contain relationships between users who comment on questions on a mathematics website (\url{MathOverflow.com}). The hypergraphs consist of $N=423$ and $N=580$ nodes, respectively. The interactions between users are mainly about comments, questions, and answers to algebra (or geometry) problems. Each node represents a user, and users who commented, asked, or answered the same question (in the area of algebra or geometry) are included in the same hyperedge. 

\textit{Restaurants-Rev}~\cite{amburg2020fair, amburg2022diverse} shows the restaurant review, which consists of $N=565$ nodes. Each node and hyperedge represent a user and the set of users who reviewed a certain restaurant, respectively. 

\textit{Music-Rev}~\cite{ni2019justifying} ($N=1106$) represents the music review of users on Amazon. Nodes and hyperedges denote users and a set of reviewers who reviewed a certain category of blues music over a one-month period.

\textit{NDC-classes}~\cite{Benson-2018-simplicial, yoon2020much} is a dataset containing drug information. It contains $N=1161$ nodes. The data characterize the association between drugs in terms of their class labels. Note that drugs with the same label are included in one hyperedge. 

\textit{Bars-Rev}~\cite{amburg2020fair} ($N=1234$) indicates a bar review hypergraph, where each node denotes a Yelp user and each of the hyperedges indicates a set of users who reviewed a specific type of bars in Las Vegas in a month's time. 

Two metabolic hypergraphs in biological studies, i.e., \textit{iAF1260b}~\cite{king2016bigg} and \textit{iJO1366}~\cite{king2016bigg}, are also considered. In these data, a node represents a reaction-based metabolism, and a hyperedge is a collection of metabolic applied to a particular reaction. The two hypergraphs consist of $N=1668$ and $N=1805$ nodes, respectively. Note that the duplicate hyperedges are removed.

We investigate the topological properties of the empirical hypergraphs above, e.g., the number of nodes and hyperedges, the average degree of nodes, the average hyperdegree of nodes, the average size of hyperedges, the average length of the shortest path, the diameter and the link density, which are shown in Table~\ref{tab:topo}. 

\bibliography{sample}  

\noindent\textbf{Acknowledgments}\\
This work was supported by the National Natural Science Foundation of China (Grant No. 72371224), the Natural Science Foundation of Zhejiang Province (Grant Nos. LQ22F030008), the Major Project of The National Social Science Fund of China (Grant No. 19ZDA324), the Scientific Research Foundation for Scholars of HZNU (2021QDL030), and the Fundamental Research Funds for the Central Universities.\\

\noindent\textbf{Author contributions}\\
M.X., X.-X.Z., C.L., Z.-K.Z. designed the study; M.X., X.-X.Z., C.L., Z.-K.Z. performed the experiments and analyzed the results; M.X. and X.-X.Z. wrote
the first draft.  M.X., X.-X.Z., C.L., Z.-K.Z. contributed to the current draft.\\

\noindent\textbf{Competing interests}\\
The authors declare no competing interests.\\

\noindent\textbf{Supplementary information}\\
The supplementary material is available at the file of Supplementary Information.

\end{document}